\documentclass[superscriptaddress,reprint,amsmath,amssymb,nofootinbib,
 aps]{revtex4-1}

\usepackage{graphicx,subfigure}
\usepackage{dcolumn}
\usepackage{bm}
\usepackage{hyperref}
\usepackage[mathlines]{lineno}
\graphicspath{{Figures/}}
\newcommand{\mt}[1]{\textrm{\tiny #1}}
\newcommand{\be}{\begin{equation}}
\newcommand{\ee}{\end{equation}}
\newcommand{\bea}{\begin{eqnarray}}
\newcommand{\eea}{\end{eqnarray}}
\newcommand{\rh}{r_\mt{H}}

\begin{document}
\title{ Determination of dynamical exponents of graphene at\\ quantum critical point by holography} 

\author{Geunho Song} 
\affiliation{Department of Physics, Hanyang University, Seoul 04763, Korea.}
\author{Yunseok Seo}
\affiliation{School of Physics and Chemistry, Gwangju Institute of Science and Technology, Gwangju 61005 Korea}
\author{Sang-Jin Sin}
\affiliation{Department of Physics, Hanyang University, Seoul 04763, Korea.}

\date{\today}

 \begin{abstract}   
We calculate   the transport of a theory with two conserved currents by the holographic method and compare it with graphene data to  determine its dynamical exponents $(z,\theta)$, which characterizes a ``quantum critical point''.  
 As a result, we find that the electric and the thermal conductivity  data   can be fit more naturally  if we assume  $(z,\theta)=(3/2,1)$ rather than (1,0). 
Furthermore, we find that thermoelectric power data at high temperature can be fit if we use (3/2,1) but not by (1,0). 
The $\theta=1$ result can be interpreted as taking into account the  fermionic nature of the electrons  and $z=3/2$ can be interpreted as  the flattened  band by the strong interaction.
\end{abstract}
 \maketitle
\section{Introduction} 
The strong correlation is the property of a phase of general matter because even a weakly interacting material can become strongly interacting in some parameter region. It happens when the  Fermi surface (FS) is tuned to be small or when the conduction band is designed to  be flat. The Coulomb  interaction in a metal is small  only  because the charge is screened by the particle-hole pairs, which are abundantly created when the FS is large.   In fact, any Dirac material  is strongly correlated as far as its FS is near the tip of the Dirac cone.  This was demonstrated in clean graphene  \cite{pkim,Lucas:2015sya}  and  the surface of  topological insulator \cite{liu2012crossover,zhang2012interplay,bao2013quantum}  through the anomalous   transports   that could  be quantitatively explained by a holographic theory\cite{Seo:2016vks,Seo:2017oyh,Seo:2017yux}.  
In twisted bilayered graphene\cite{cao2018unconventional,cao2018correlated}, flat band appears due to the formation of effective lattice system called Moire lattice, which has larger size than the original lattice.  
In short,  strong correlation phenomena are ubiquitous,  where  the traditional methods are not working very well; therefore, a new method   has been longed for for many decades.   

The  strongly interacting system(SIS) is hard to be characterized   in terms of its basic building blocks and one faces the question how to simplify the system to make a sensible physics with only a few parameters. 
One possibility is that  they become simple at the quantum critical point (QCP) by the universality coming from the loss of system information, which  is similar to a black hole system.
 In this sense, the SIS and holographic theory are similar by sharing the property of black holes.  
A  QCP is characterized by $z, \theta$ defined by the dispersion relation of excitations 
$\omega\sim k^{z}$  and the entropy density $s\sim T^{(d-\theta)/z}$. Interstingly,  there exists a  metric with the same scaling symmetry   $(t,r,x)\rightarrow(\lambda^zt,\lambda^{-1}r, \lambda x)$,
\bea
	ds^{2}=r^{-\theta}\bigg(-r^{2z}dt^{2}+\frac{dr^{2}}{r^2 }+r^{2}(dx^{2}+dy^{2})\bigg),
\eea
 which is called the hyperscaling violation (HSV) metric. 
 
The purpose of this paper is to reexamine the transport data of graphene  
  to  determine its dynamical exponents $(z,\theta)$, which characterizes a QCP.      In our previous work \cite{Seo:2016vks},  we  assumed that  the theory has a QCP at $(z,\theta)=(1,0)$ based on the presence of the Dirac cone, and showed that there must be at least two   conserved currents. We also had to assume that the entropy density is a free parameter to be tuned to fit the data.
  
 In this paper, we extend 
  the    holographic theory with two currents 
\cite{Seo:2016vks,Rogatko2018} using the HSV geometry. 
As a result, we could eliminate  the last  assumption, namely,  the entropy density is not assumed to be a free parameter but a physical quantity  determined by other parameters, which is a progress.  
The electric and the thermal conductivity  data   can be fitted much more naturally  if we assume  $(z,\theta)=(3/2,1)$ rather than (1,0).  We also find that thermoelectric power data at high temperature can be fitted if we use (3/2,1) but not at all by (1,0).
Our work demonstrates  that   critical  exponents  together with the ratio of the  conserved charges   completely determines the transport data of a strong correlated system.

Notice that our   system has  $z=1$ at UV.  But nevertheless, due to the interaction between the gravity and the   spin 1 field $A_{\mu}$,  the system can have a solution with $z>1$ at IR.  Therefore conceptually,  it is good idea to assume that HSV is embedded into an asymptotically AdS solution.   
Recently, we noticed that Sungsik Lee and Metlistki+Sachdev \cite{PhysRevB.82.075127, PhysRevB.78.085129}, using  renormalization group method, 
 found that due to the curvature effect of the Fermi surface, the fermion dispersion relation near the IR fixed point has z=3/2, although a physical situation was not targeted as graphene but the spin liquid. However, their formulation seems to be universal. They   assumed just the presence of a Fermi surface and Yukawa interaction.  
The fact that  we can get a consistent result   by solving a ``classical equation'' is certainly a great power of the holography, and it is a true evidence that holographic theory encodes the quantum information.

 \section { hyperscaling violating geometry with two currents }  
 We   consider a four-dimensional action with an asymptotically AdS  metric $g_{\mu\nu}$, a dilaton field $\phi$,  three  spin 1 fields $A_{\mu}, B^{1}_{\mu},B^{2}_{\mu}$. We also use two scalar fields $\chi_1,$ $\chi_2$ called axions to break the translational symmetry. The action is given by $S=\int_{\mathcal{M}}d^4x\mathcal{L}$, with 
 \bea\label{action1}
\mathcal{L}=\sqrt{-g}\bigg(R&+& {\cal V}e^{\gamma\phi}-\frac{1}{2}(\partial\phi)^2-\frac{1}{4}Z_A F^2\nonumber
\\&-&\sum_a^2\frac{1}{4}Z_a G_{(a)}^2-\frac{1}{2}Y\sum\limits_{i}^{2}(\partial \chi_i)^2\bigg)
\eea
where $F=dA$, $G_{(a)}=dB_a$. 
We use an ansatz,  
\be
Z_A= e^{\lambda \phi}, \quad Z_a=\bar{Z}_ae^{\eta\phi}, \quad Y=e^{-\eta\phi},\quad \chi_i=\beta x_i,
\ee
where $\eta, \lambda,$ and $\nu$ are dimensionless  numbers and  $\beta$ denotes the strength of momentum relaxation and can be interpreted as the density of impurity. 
Notice that 
 there are three gauge fields  in our model. 
$A_{\mu}$ is a part of gravity. It should be considered as a Kaluza-Klein reduction of a higher dimensional metric.  Indeed its existence is solely to support the Lifshitz gravity with HSV. 
The other two spin 1 fields are gauge  fields that are dual to two independent currents. 
In Ref. \cite{Seo:2016vks}, these two currents were introduced  
because there are two independent currents due to the imbalance effect. Briefly, 
the electron current and the hole current are separately conserved  because the process 
$e \to e+e+h$, which is necessary to balance the deficit electron, and a similar process to create more holes are forbidden in the small timescale since the energy-momentum conservation requests collinearity of the four momenta, which  is a zero subset measure of the phase space. For more detail,   see  \cite{Seo:2016vks,Foster}. Introducing two currents is also necessary because with single current, one can not fit the data quantitatively. 
 
 The equations of motion of   fields  are 
 \begin{eqnarray}
&&\partial_{\mu}(\sqrt{-g}Z_AF^{\mu\nu})=0, \quad 
\partial_{\mu}(\sqrt{-g}Z_aG^{\mu\nu}_{(a)})=0,\label{max}  \\
&&R_{\mu\nu}-\frac{1}{2\sqrt{-g}}g_{\mu\nu}\mathcal{L}-\frac{1}{2}\partial_{\mu}\phi\partial_{\nu}\phi-\frac{Y}{2}\sum_i \partial_{\mu}\chi_i\partial_{\nu}\chi_i    \nonumber\\
&&\quad\quad\quad\quad\quad-\frac{1}{2}Z_A F_{\mu}^{\rho}F_{\nu \rho}-\sum_a^2\frac{1}{2}Z_a G_{(a)\mu}^{\rho}G_{\nu \rho}^{(a)} \label{eins} =0,\\
&& \Box\phi+  {\cal V}\gamma e^{\gamma\phi}-\frac{1}{4} Z_A'(\phi) F^2-\frac{1}{4}\sum_a Z_a'(\phi) G^2_{(a)}\nonumber\\
&&\quad\quad\quad\quad\quad\quad-\frac{1}{2}Y'(\phi)\sum\limits_{i=1}^{2}(\partial \chi_i)^2=0,\\
&&\partial_{\mu}(\sqrt{-g}g^{\mu\nu}Y\sum_{i=1}\partial_{\nu}\chi_i)=0.  
\end{eqnarray}
The solutions for the  fields are given by 
\be
\phi(r)=  \nu\ln r, \quad \hbox{with } \nu=\sqrt{(2-\theta)(2z-2-\theta)}.
\ee
\begin{eqnarray}\label{bgsol}
&&\quad A=a(r)dt, ~B_1=b_1(r)dt,~ B_2=b_2(r)dt,   \\
&&\quad \chi=(\beta x, \beta y),  \\
&&ds^{2}=r^{-\theta}\bigg(-r^{2z}f(r)dt^{2}+\frac{dr^{2}}{r^2 f(r)}+r^{2}d\vec{x}^{2}\bigg), \\
&&f(r)=1-m r^{\theta-z-2}-\frac{\beta^{2}}{(\theta-2)(z-2)}r^{\theta-2z}\nonumber\\
&&\qquad\qquad\qquad+\frac{(Z_1q_1^{2}+Z_2q_2^2)(\theta-z)r^{2\theta-2z-2}}{2(\theta-2)}\\
&&a(r)=\frac{-q_A}{2+z-\theta}(\rh^{2+z-\theta}-r^{2+z-\theta}), \nonumber\\
&& b_a(r)= \big(\mu_a-q_a r^{\theta-z} \big), 
\end{eqnarray}
where $a=1,2$. Here, all the coordinate and parameters are rescaled by the anti-de Sitter (AdS) scale $L$  to make them  dimensionless. 
 We  set $L=1$ 
until the last moment to avoid introducing a dimensionless version of the parameters. 
 The gauge couplings $Z_1$ and $Z_2$ are then given by  
\bea
Z_A(\phi)=e^{\lambda\phi},\quad Z_a(\phi)=\bar{Z}_ae^{\eta\phi}, \quad  Y(\phi)=e^{-\eta\phi}
\eea
with
\bea
&&\lambda=\frac{\theta-4}{\nu},\;  \eta=\frac{\nu }{2-\theta}, \; 
\gamma=\frac{\theta}{\nu }, \; {\cal V}=\frac{z-\theta+1}{2(z-1)}q_A^2, \\
&&q_A=\sqrt{(2z-2)(2+z-\theta)}.   
\eea
There are important restrictions in the range of the 
parameters for the HSV solution coming from the null energy condition and the positivity of $q_{A}^{2}$, which is studied in the previous work\cite{Ge:2019fnj}, and we attached it in the Appendix. 
The presence  of the singularity in the HSV  geometry was pointed out in \cite{Lei2013} and it was shown that it can be resolved if $\nu=0$ and $1\le\theta\le2$. We can check that the   the screened HSV geometry are always regular independent of $(z,\theta)$. See the Appendix.
\begin{figure}
\begin{center}
\subfigure[Quantum Critical Region(QCR)]{\includegraphics[angle=0,width=0.4\textwidth]{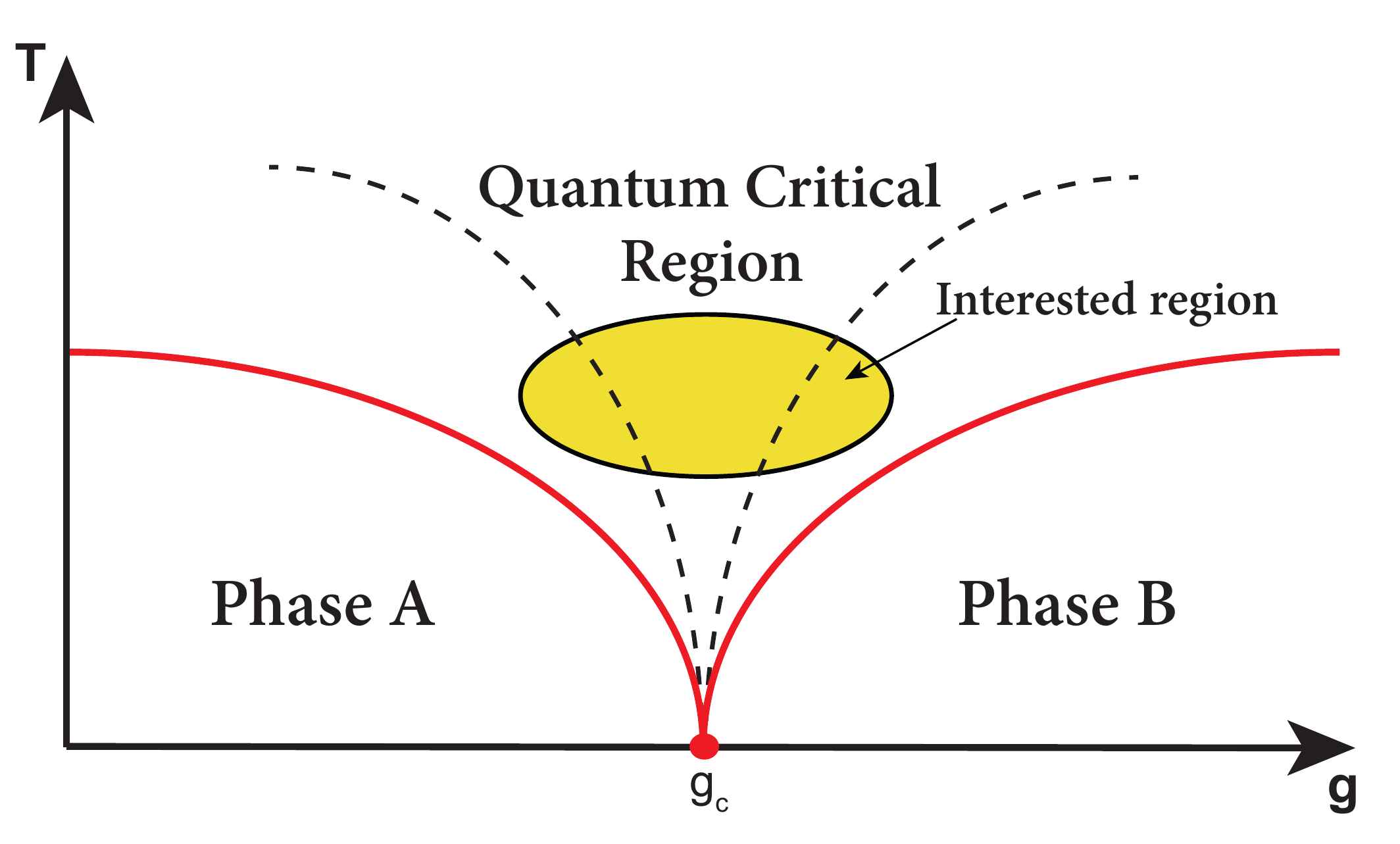}}
\subfigure[The scheme of embedding HSV to AdS]{\includegraphics[angle=0,width=0.5\textwidth]{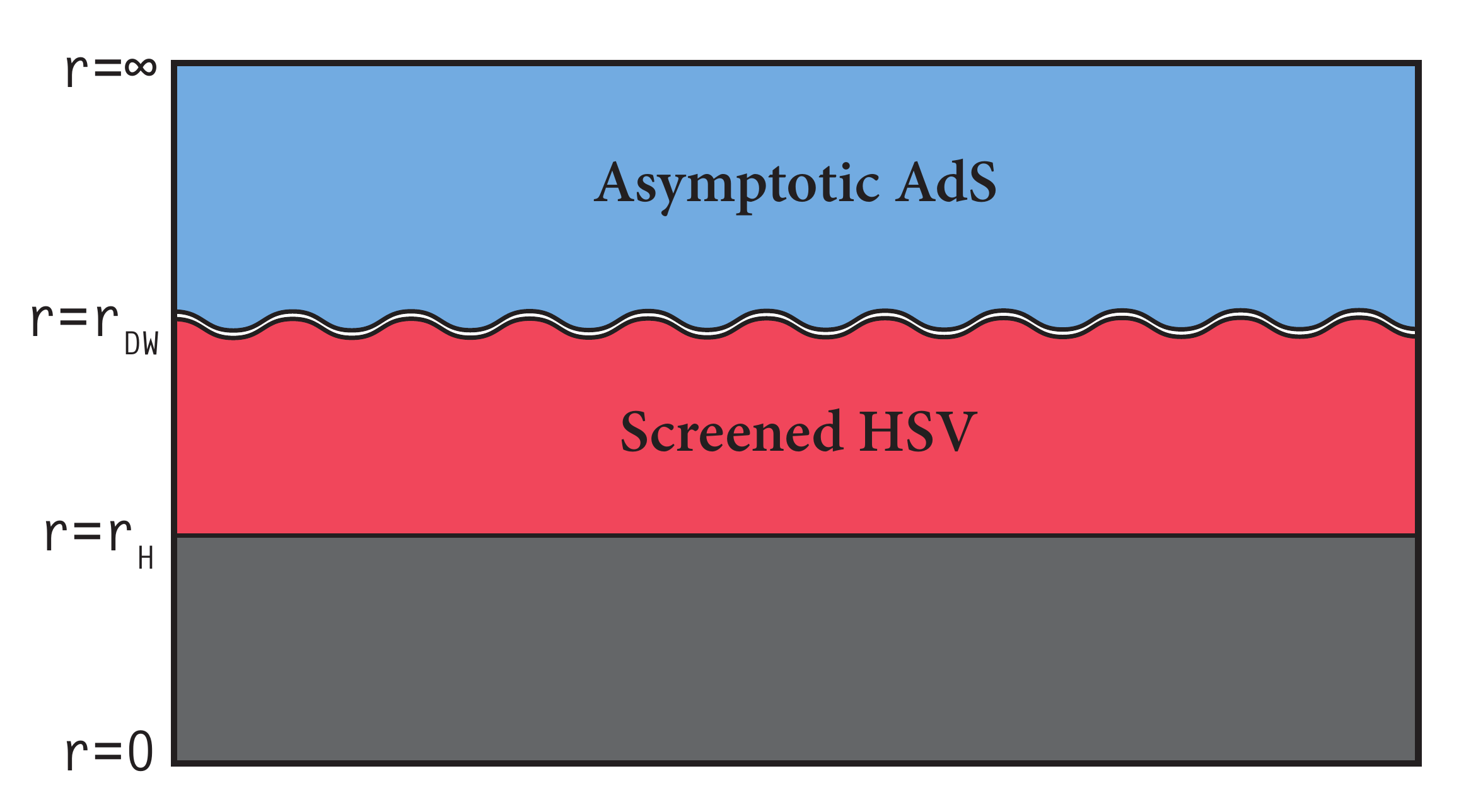}}
\caption{(a) Phase diagram with quantum critical point which is denoted by $g_c$ where $T=0$. But we are interested in the finite temperature region that can be affected by the quantum critical point, which is called quantum critical region (QCR). Here, we consider the region colored by yellow. (b) The schematic figure of our geometry. The region inside the black hole is colored with gray, and there is a domain wall $(r=r_{DW})$ somewhere between the $AdS$ boundary and black hole horizon at $r_{H}$. 
   } \label{region} 
\end{center}
\end{figure}

Notice that $a(r)$ is divergent.  From the early time of Lifshitz gravity, such a divergence of the $A_{\mu}$ has been controversial. 
At this moment, the   consensus is that HSV geometry should be embedded into asymptotically AdS spacetime so that it is just the IR part of the total domain-wall solution. 
See Fig. \ref{region}(b).
A related question is about the scaling property of $A_{\mu}$. At zero temperature and zero densities ($B_{1\mu},B_{2\mu}=0$), $A_{t}$ has the  scaling properties $A_{t}\to \lambda^{2+z-\theta}A_{t}$. If any of these quantities are nonzero, there is no scaling property. 
However, notice that we are not describing a quantum critical point itself, which is at zero temperature. Our interest is the quantum critical region (QCR) above that point, where scale symmetries of all the gauge field are   broken by the existence of the temperature and  chemical potential for $B_{1\mu},B_{2\mu}$.   See Fig. \ref{region}(a).

From the equations of motion for the gauge fields $B$'s,  
 we can obtain the charge density as the integration constants,
\bea
	Q_a=\sqrt{-g}Z_aG_{(a)}^{tr}=\bar{Z}_a q_a (z-\theta). 
\eea
The entropy density and the Hawking temperature  are
\bea
s=&&4\pi r^{2-\theta}_H, \\
{4\pi}T=&& (z+2-\theta)\rh^z -\frac{\beta^2 \rh^{\theta-z}}{2-\theta}\nonumber\\
&&\qquad-\frac12\left(\frac{Q_1^2}{\bar{Z}_1}+\frac{Q^2_2}{\bar{Z}_2}\right) \frac{\rh^{2\theta-2-z}(z-\theta)^2}{(2-\theta)} .
\label{r0Trelation}
\eea

\section{Calculation of DC transport}
We use  following perturbation to compute the transport coefficients \cite{Donos:2014cya}:
\bea
	&&\delta g_{tx}=h_{tx}(r)+tf_{3x}(r),~~ \delta g_{rx}=h_{rx}(r), \nonumber\\
	 &&\delta B_{ax}=b_{ax}-tf_{ax},~~ \delta\chi_1=\varphi_{x}(r)
\eea
where   linearized $f_{i}$'s are chosen, such that they provide 
time-independent source terms in Einstein equations,
\bea
f_{ax}=-E_a+\zeta b_a(r), \quad
 f_{3x}=-\zeta U(r)
\eea
where $a=1,2$. Here, $E_1, E_2$ are electric forces acting on $J_1, J_2$, respectively, and $\zeta$ is a thermoelectric force given by the temperature gradient $\zeta=-(\nabla T/T)$. We will set $E_{1}=E_{2}=E$, after the calculation is done.  The transports can be computed at the event horizon using the Maxwell equation that provides the conservation of the currents in a radial direction. 
 In the Eddington-Finkelstein (EF) coordinates $v,r$  the background metric is given by 
 \bea
 	ds^2&&=-Udt^2 +{V}dr^{2}+Wd\vec{x}^2,\nonumber\\
	&&=-Udt^2-2\sqrt{UV}dvdr+Wd\vec{x}^2,
\eea
with $v=t+\int dr\sqrt{V/U}$. Notice that  it is  regular at the horizon.  If we turn on $\delta g_{rx}=h_{rx}$ and $\delta g_{tx}=h_{tx}$,   the   metric   perturbation in  EF coordinates, can be written as 
\bea
	\delta g_{\mu\nu}dx^{\mu}dx^{\nu}=h_{tx}dvdx+\left( h_{rx}-\sqrt{\frac{V}{U}}h_{tx}\right)drdx.\nonumber
\eea
To guarantee its regularity at the horizon,   the last term is requested  to vanish at the horizon so that 
\be 
h_{rx}\sim\sqrt{\frac{V}{U}}h_{tx}.\label{reg1} \ee 
Similarly, we can express the gauge field perturbation  in the EF coordinates as 
\bea \label{horizon1}
	\delta B_{ax}\sim b_{ax}+E_av-E_a\int dr\sqrt{\frac{V}{U}}. 
\eea
One should note that $b_{a}$ is the background solution of the time component gauge field and $b_{ax}$ are fluctuations of gauge field components $B_{ax}$. 
Then, gauge field perturbation  will take the regular form $\delta B_{ax}\sim E_a v+\cdots$ by  demanding 
\bea
	b_{ax}'\sim\sqrt{\frac{V}{U}}E_a, \quad a=1,2 .\label{reg2} 
\eea
Now, the $rx$ component of the Einstein equation is 
\bea
	\frac{Y\beta^2 }{W}h_{rx}-\frac{1}{U}\left(\sum_{a=1,2}(Z_a b_a'f_{ax})+f_{3x}'\right)+\frac{W'f_{3x}}{UW}=0.\nonumber
\eea
For the regularity   at the horizon,  we request 
\bea \label{horizon2}
	h_{tx}|_{r_H}=-\frac{1}{\beta^2 Y}\left(s T \zeta +\sum_{i=1,2}\bar{Z}_iq_iE_i(z-\theta)\right) \label{reg3}
\eea
where we used Eq. (\ref{reg1}). 
 The Eqs (\ref{reg1}), (\ref{reg3}), and eq.(\ref{reg2}) are the  regularity conditions for the metric and the  gauge fields at the event horizon. 
The Maxwell equations (\ref{max}) give  conserved currents  \cite{Donos:2014cya},
\bea
	J_a&=&\sqrt{-g}Z_aG_{(a)}^{xr},\\
	\mathcal{Q}&=&\frac{U^2}{\sqrt{UV}}\left(\frac{h_{tx}}{U}\right)'-\sum_{a=1,2}b_a J_a, 
\eea
where the index $a=1,2$ is for  two currents, which are dual to the two gauge fields $B_{a}$. 
These currents are radially conserved so that their boundary values can be computed at the horizon \cite{Donos:2014cya}. Therefore,  we can get the boundary current in terms of their horizon behavior (\ref{horizon1}), (\ref{horizon2}), which again is given by the external sources,
\bea \label{currentsources}
	J_1&=&\left(Z_1+\frac{Q_1^2}{WY\beta^2}\right)E_1+\frac{Q_1 Q_2}{WY\beta^2}E_2+\frac{4\pi T Q_1}{Y\beta^2}\zeta\nonumber\\
        J_2&=&\frac{Q_1 Q_2}{WY\beta^2}E_1+\left(Z_2+\frac{Q_2^2}{WY\beta^2}\right)E_2+\frac{4\pi T Q_2}{Y\beta^2}\zeta\nonumber\\
	\mathcal{Q}&=&\frac{4\pi T Q_1}{Y\beta^2}E_1+\frac{4\pi T Q_2}{Y\beta^2}E_2+\frac{16\pi^2 W T^2}{Y \beta^2}\zeta
\eea
We can write (\ref{currentsources}) in matrix form, $J_i=\Sigma_{ij}E_j$, with $J_3=\mathcal{Q}$ and $E_3=\zeta$,
\begin{align}\label{trans_matrix}
\left(\begin{array}{ccc}     
 \sigma_{11} &\sigma_{12}  &\alpha_1 T  \\  
\sigma_{21} &\sigma_{22}&\alpha_2 T \\  
\bar{\alpha}_1 T & \bar{\alpha}_2 T&\bar{\kappa} T 
\end{array}\right) :=\Sigma .
\end{align} 
Notice that the matrix $\Sigma$ is symmetric so that 
\bea
	\sigma_{12}=\sigma_{21},\qquad \alpha_i=\bar{\alpha}_i
\eea
The heat conductivity $\kappa$ is defined by the response of 
the  heat current to the temperature gradient $\zeta$ in the absence of electric currents $J_1$ and $J_2$: we can express $E_1$ and $E_2$ in terms of $\zeta$ by setting $J_1$ and $J_2$ to vanish in (\ref{currentsources}). Substituting these expression for $E_i$ to the last line of (\ref{currentsources}) and taking   derivative with respect to the temperature gradient, we can get
\begin{align}\label{thermal_cond}
\kappa&=\bar{\kappa} - \frac{T\bar{\alpha}_1( \alpha_1 \sigma_{22} - \alpha_2 \sigma_{12})}{ \sigma_{11} \sigma_{22} -\sigma_{12}\sigma_{21}} -  \frac{T\bar{\alpha}_2(\alpha_2 \sigma_{11} - \alpha_1 \sigma_{21})}{  \sigma_{11} \sigma_{22} -\sigma_{12} \sigma_{21} }
\end{align}
with $\bar{\kappa}=4\pi s T/Y\beta^2$. The Seebeck coefficient is defined by 
\bea
	S_i=\sum_{j}\sigma^{-1}_{ij}\alpha_j
\eea
Then, the transport coefficients for conserved currents can be calculated as the following:
\bea
	\sigma_{ij}&=&Z_i\delta_{ij}+\frac{Q_iQ_j}{WY\beta^2},\quad \alpha_i = \frac{4\pi Q_i}{Y\beta^2}, \quad \bar{\kappa}=\frac{16\pi^2 WT}{Y\beta^2}\nonumber\\
	 \kappa&=&\frac{\bar{\kappa}}{1+\sum_i4\pi Q_i^2/sZ_i Y\beta^2}\nonumber\\
	 S_i&=&\frac{sQ_i/Z_i}{WY\beta^2+\sum_i(Q_i^2/Z_i)}  
\eea
If we define the total electric current as $J=\sum_iJ_i$ and the thermoelectric force as $E_i=E-T\nabla(\mu_i/T)$, the electric conductivity  based on total current is given by
\bea
	\sigma=\frac{\partial J}{\partial E}=\sum_{ij}\sigma_{ij}=Z+\frac{Q^2}{WY \beta^2}
\eea
{where $Q=\sum_i Q_i$ and $Z=\sum_i Z_i$, showing the additivity of the charge-conjugation-invariant part \cite{Blake:2014yla} of the electric conductivity. If we define   the heat conductivity due to the $i$th current by $1/\kappa_{i}=1/{\bar\kappa}+Q^{2}_{i}/Z_{i}s^{2}T$,   
then  the heat conductivity formula leads us to {\it the additivity of the dissipative part of the inverse heat conductivity}. Therefore, 
\be 
D[1/\kappa]=\sum_{i}D[1/\kappa_{i}],\quad \quad
{\bar D}[ \sigma]=\sum_{i}{\bar D}[\sigma_{i}],
\label{additivity}\ee 
where  $D[f]$ denotes the dissipative  part of $f$ and ${\bar D}[f]=f-D[f]$ .}

 The total Seebeck coefficient $S$  by the two currents is given by 
 \bea
 	S=S_{1}+S_{2}
	=\frac{4\pi W(Z_1 Q_2+Z_2 Q_1)}{Z_1 Q_1^2+Z_2 Q_2^2+Z_1Z_2WY\beta^2}
 \eea


Finally,   the two currents are independently conserved for short moments but   long enough for the hydrodynamic equilibrium to be  reached,  as argued in 
 \cite{Seo:2016vks,Foster}. 
In this case, individual charges, the hole and electron charges, are separately conserved, 
therefore,
 \begin{align} 
Q_i=g_{i} Q, \label{qlinearity}
\end{align} 
for some $g_{1},g_{2}$.  
Then the experimental data of graphene will be well fit by our  two current theories, as we will see below.  

\section{ Theory vs experiments}
\subsection{Thermoelectric power}
\begin{figure}
\begin{center}
\subfigure[Electric conductivity]{\includegraphics[angle=0,width=0.4\textwidth]{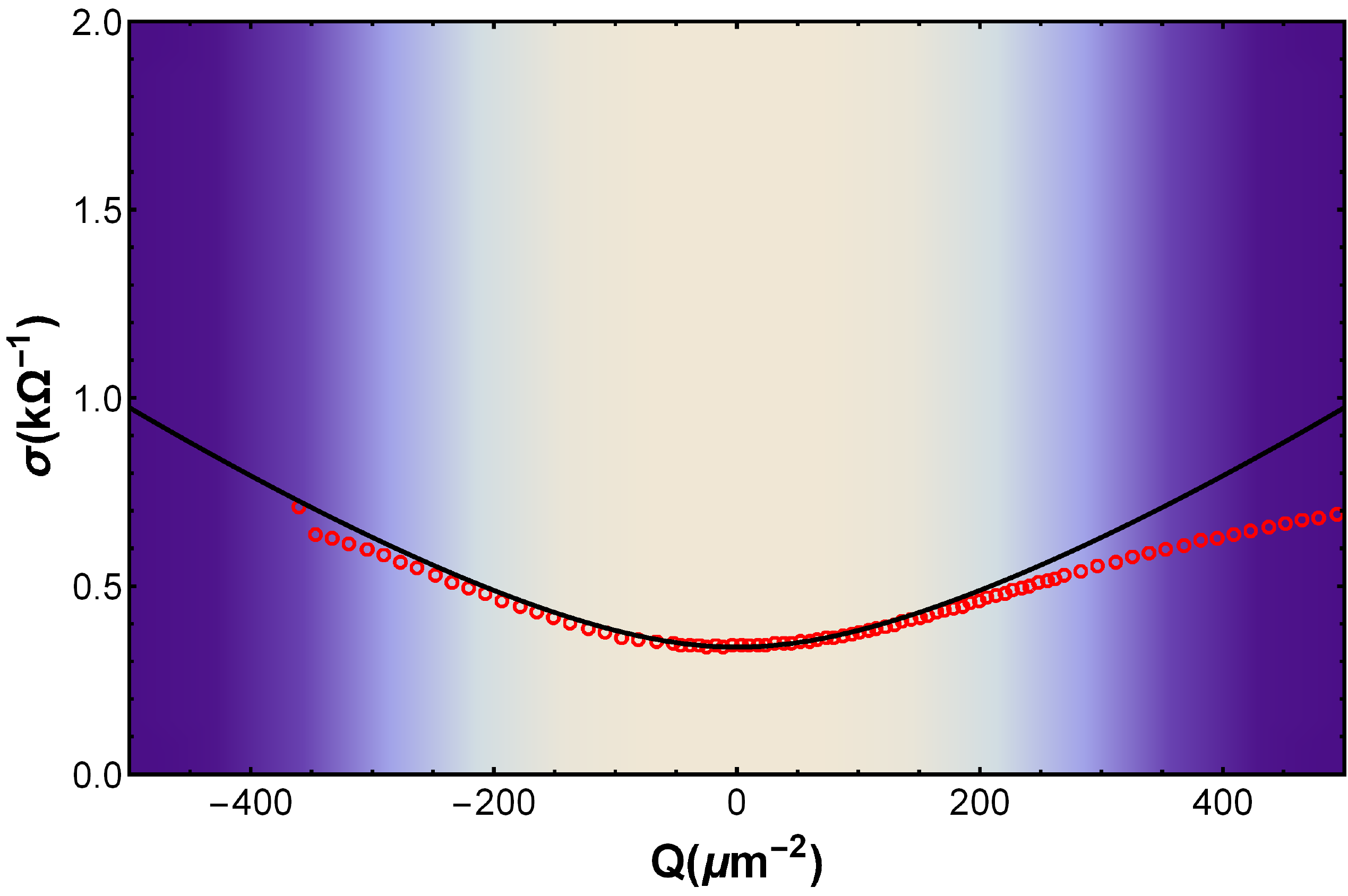}}
\subfigure[Thermal conductivity]{\includegraphics[angle=0,width=0.4\textwidth]{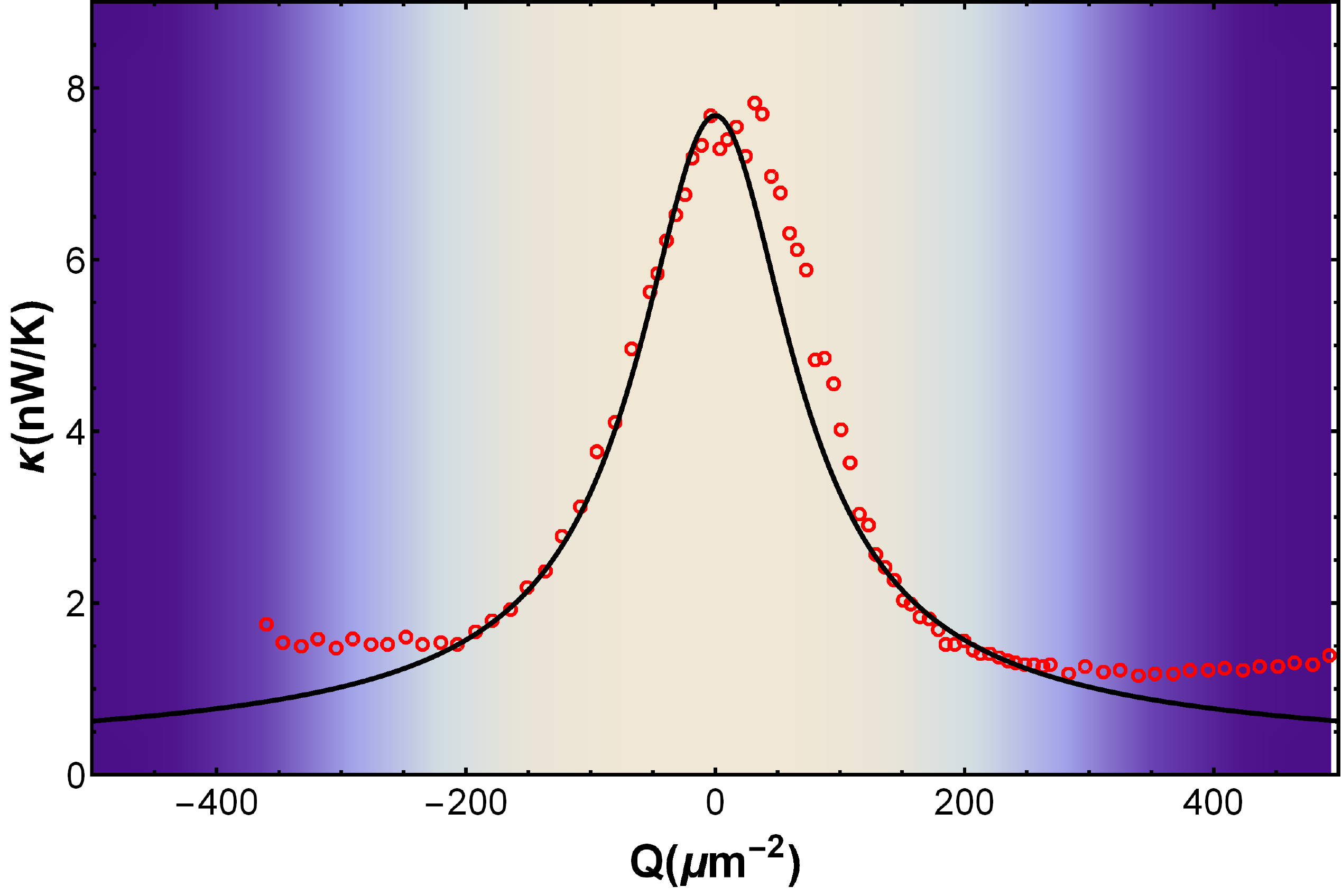}}
\caption{Comparison with real experiment : (a) density plot of electric conductivity $\sigma$ and (b) of thermal conductivity $\kappa$. Red circles are for data used in \cite{pkim,Lucas:2015sya}, and black curves are for two current model. The region shaded with blue is for the Fermi liquid which is far from our theory. 
   } \label{fig} 
\end{center}
\end{figure}

  The total electric current $J$ and total number current $J_n$ are defined by $J=J_e+J_h$, $J_n=J_e-J_h$, respectively, and their corresponding densities (electric charge densities and number densities) are related by $Q_1=q_en_1$ and $Q_2=-q_en_2$ with a charge of an electron $q_e=-1$. The total electric charge density and total number density are defined by $Q=Q_1+Q_2$ and $Q_n=-Q_1+Q_2$, which can be connected with the proportionality constant $g_n$ such that $Q_n=g_nQ$. 
  Notice that $\eta=0$ when $z=(\theta+2)/2$ so that $Z_a=\bar{Z_a}$ and $Y=1$. From now on, we take $z=3/2$, $\theta=1$. There are two reasons for choosing this $(z,\theta)$: (i) $\theta=1$ is necessary to encode the fermionic nature of the system. (ii) $z=3/2$ is the optimized dynamical exponent for fitting the experimental results which will be shown later. Then, the total electric conductivity $\sigma=\frac{\partial J}{\partial E}$ and $\kappa$ can be expressed in terms of $Q$ and $g_n$,
    \bea
  	\sigma=\sigma_0\left(1+\frac{Q^2}{Q_0^2}\right), \quad \kappa=\frac{\bar{\kappa}}{1+(1+g_n^2)(Q/Q_0)^2}.
  \eea
 where $\sigma_0=2Z_0$ and $Q_0= {\sigma_0 s\beta^2}/{4\pi}$.
  
Notice that in all our formula so far, we used dimensionless version of the parameters, which was 
introduced at the level of the equation of motion before we get the solution. 
However,  for the numerical fitting, all the dimensions of the parameters should be restored to their original dimensionful version.   Following the prescription for the restoration of dimensionality  is useful, 
 \be
 \beta\to \beta L, \quad T\to \frac{k_{B}T}{\hbar v_{F}}L, \quad  s \to  sL^2,\quad Q\to QL^2 \label{dimrestore}
\ee 
where $v_F\sim c/300 =1\times 10^6m/s$, which is the Fermi velocity in graphene.
\footnote{ $L$  was introduced as an AdS radius in the original top down approach, but in our bottom up approach, it is any length scale.  
Starting from the equation of motion, we rescaled all the variables using $L$ and set it to be 1 so that every coordinate and the parameters are dimensionless. 
Notice   that the physical dimension and scaling ``dimension'' are different. 
The scaling properties are those of   dimensionless variables, and 
here, we are explaining the restoration of the physical dimension. For example,  $t$ and $x$ can have different scaling although they have the same physical dimension in a natural unit. Any quantity can be written in terms of $v_{F}^{A}\hbar^{B}L^{C}$.  
Without introducing a length scale in a theory, we can not describe or plot  any dimensionful physical quantity. 
We assumed that the Boltzman constant $k_{B}$ always follows the $T$ and $s$. }

With such a prescription, 
  \bea
  	\sigma_0=\frac{e^2}{\hbar}2Z_0, \quad \frac{\bar{\kappa}}{T}=\frac{4\pi k_B^2}{\hbar}\frac{s }{\beta^2},\quad Q_0^2=\frac{2Z_0s\beta^2 }{4\pi}L^4 . \label{restored}
  \eea
To fit the experimental results in Fig \ref{fig} for transports in graphene, we used four measured values, $\sigma_0=0.338k\Omega^{-1}$, $\bar{\kappa}=7.7nW/K$. From the curvature of the density plot of $\kappa$, we fix $g_n=3.3$ and assumed charge conjugation symmetry to set $Z_1=Z_2=Z_0$. Then, the parameters of the theory can be determined: $L = 0.2\mu m$, $2Z_0=1.387$, $\beta^2=96.75/(\mu m)^2$.  In previous work \cite{Seo:2016vks}, we replaced the horizon area  $4\pi\rh^{2}$ as the entropy density and considered the latter as a free parameter to tune.  On the other hand, we use  $\rh$ as a function of other  physical parameters such as temperature $(T)$, charge density $(Q)$, and impurity density $\beta^2$ which comes from (\ref{r0Trelation}). In that sense, we could fit the data with one less parameter. 

 \begin{figure}
\begin{center}
\subfigure[One current model]{\includegraphics[angle=0,width=0.3\textwidth]{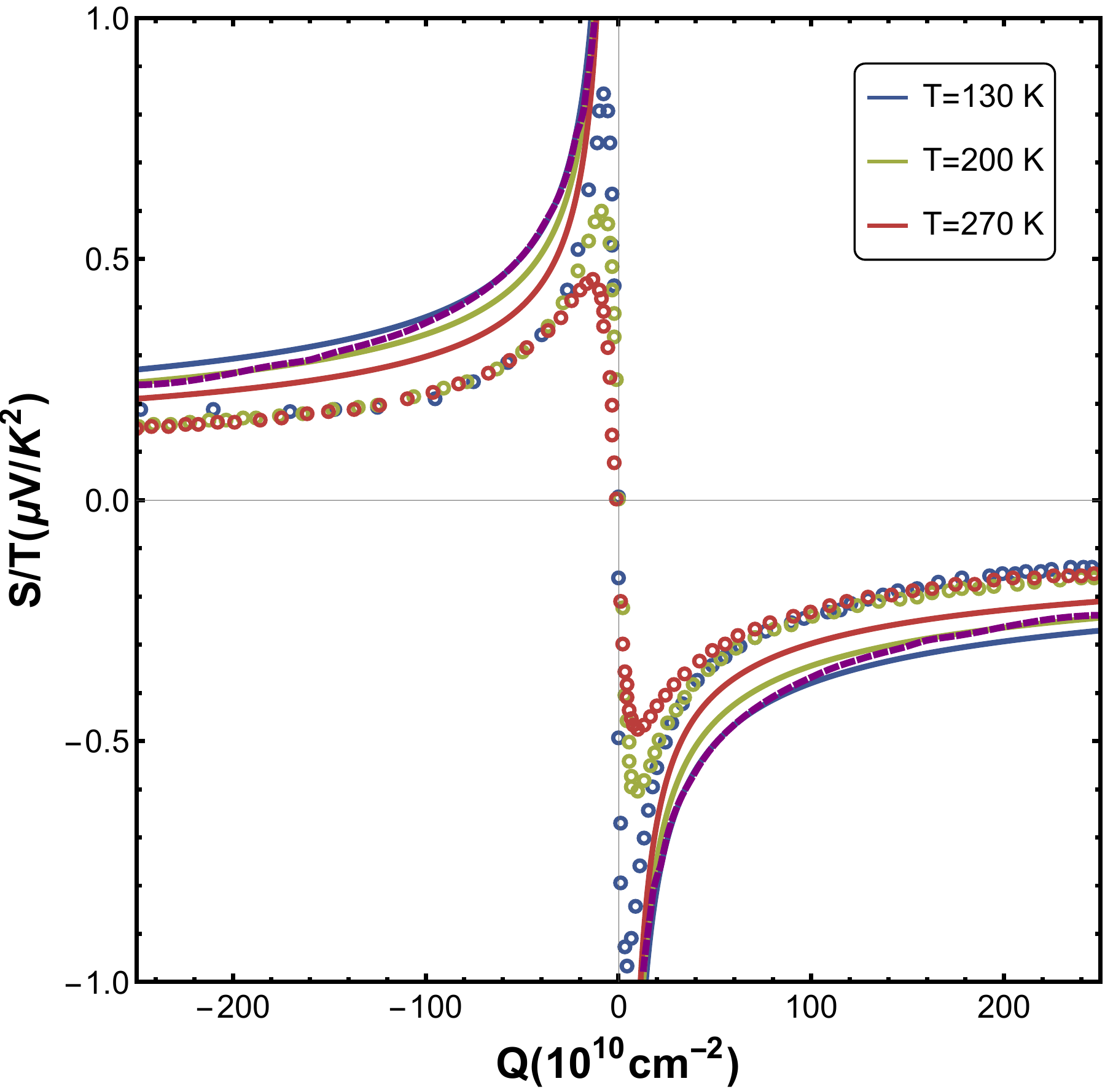}}
\subfigure[$z=1$, $\theta=0$]{\includegraphics[angle=0,width=0.3\textwidth]{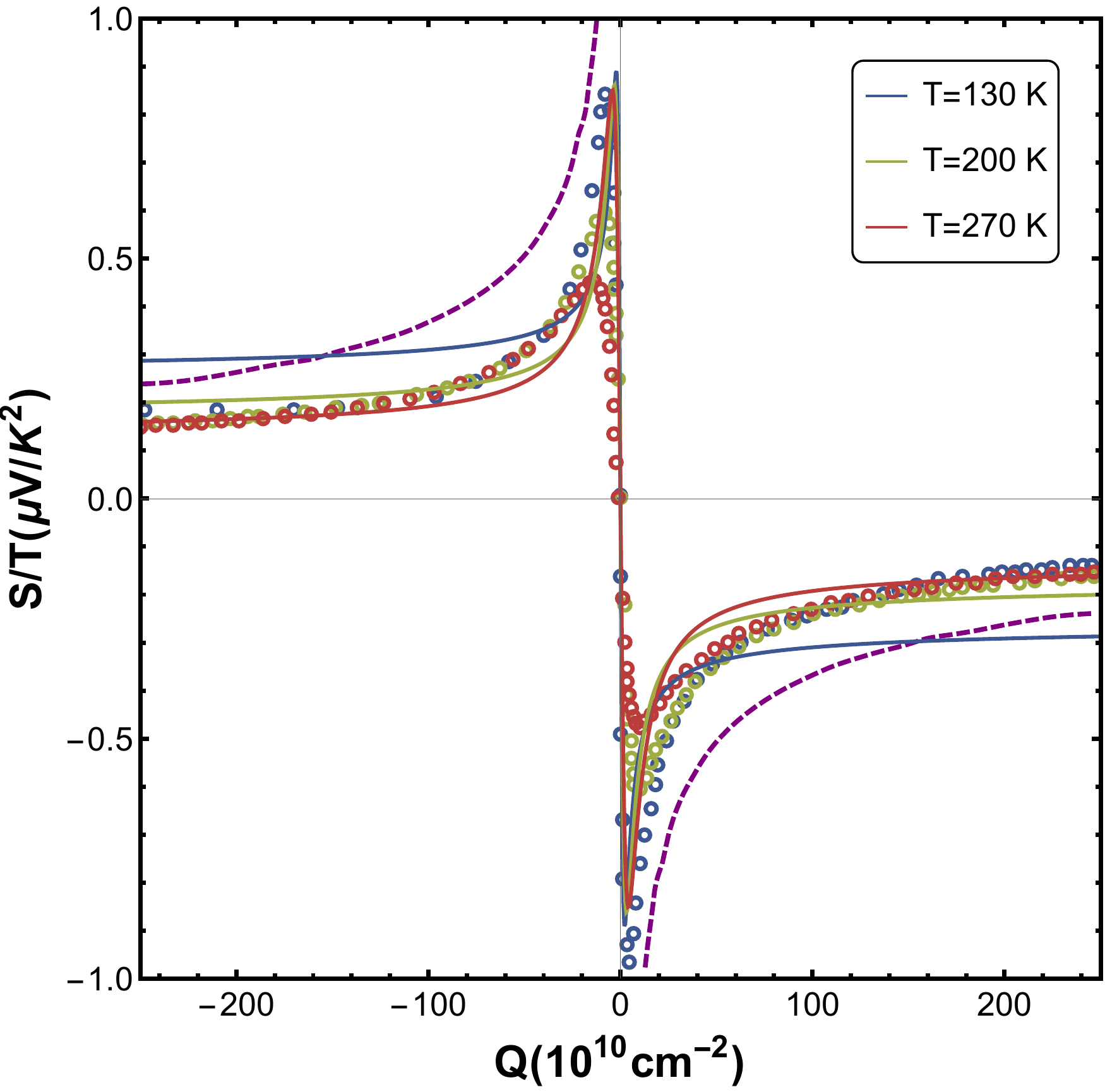}}
\subfigure[$z=3/2$, $\theta=1$]{\includegraphics[angle=0,width=0.3\textwidth]{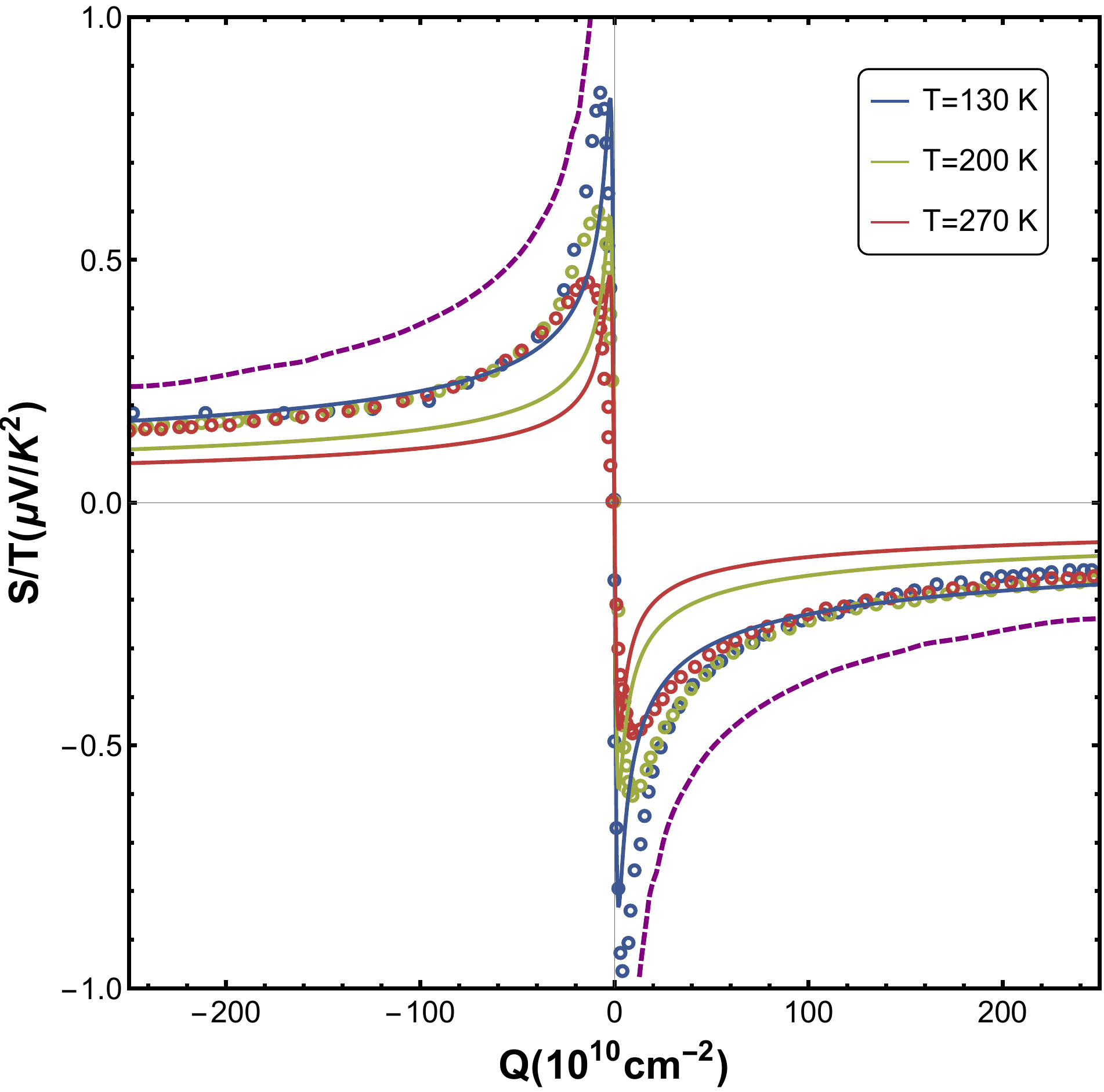}}
\caption{Theory vs data for Seebeck effect : (a)  We take $g_n=1$, which corresponds to one current model, and we set $\beta=0$ to compare with the hydrodynamics result (dashed line). (b) Seebeck coefficient for $z=1$ and $\theta=0$. We used the parameters $L = 0.2\mu m$, $\beta^2=1406/(\mu m)^2$, $2Z_0=1.387$, and $g_n=16$.   (c) For $z=3/2$ and $\theta=1$, Seebeck coefficient at low temperature fits well with experiment. We used the parameters $L=0.2\mu m$, $\beta^2=1406/(\mu m)^2$,   $2Z_0=1.387$, and $g_n=3.3$. Circles are for data used in \cite{pkim2}.
   } \label{fig2} 
\end{center}
\end{figure}

One more available data for the graphene is 
 the seebeck coefficient $S$ given in Ref. \cite{pkim2}.   $S$ can be expressed in terms of $Q$ and $g_n$:
 \bea
 	S = -\frac{k_B}{e}\frac{8\pi Q/2Z_0\beta^2}{1+(1+g_n^2)\frac{Q^2}{Q_0^2}}
 \eea
If we try to fit the experimental data of graphene,  it seems that one current model without dissipation follows the hydrodynamic model. See  Fig. \ref{fig2} (a).  For the two currents models  with $z=1$, $\theta=0$ [Fig. \ref{fig2} (b)], the theory curve of $S/T$ goes to a constant at large $|Q|$,  and the height of its peak is independent of the temperature. On the other hand, the experimental curves of $S/T$ decreasesas $|Q|$ increases, and the height of its peak also decreases as the temperature rises. Both of the features are not even close to the qualitative feature of the experimental data. 

On the other hand, for $z=3/2$ and $\theta=1$, the two currents model fits very  well with the data when $T$ is   low enough. Also, the theory curves have a tendency for convergence for large $|Q|$ and the height of its peak lowered as one raises the temperature. 
It is natural that the model does not fit with the experimental for large $T$ because our theory does not include the phonon effect, which is important for a large temperature. The reason to take $z=3/2$ is the following:  due to the Null energy condition $(2-\theta)(2z-2-\theta)\geq 0$, we cannot take $z<3/2$ with $\theta=1$, and for $z>3/2$,  the bigger value $z$ has, the greater the inconsistency of the theoretical curve with the experimental results becomes.
We conclude that the graphene data can be fit with holographic theory if we choose the dynamical exponents 
$(z,\theta)=(3/2,1)$. 

\subsection{Lorentz ratio}
In the Ref.\cite{pkim} there is a plot for the Lorentz ratio $L/L_{0}$, which could be fitted  even by an untwisted hydrodynamics using the hydrodynamic formula,
\bea
	\mathcal{L}(n)=\frac{\mathcal{L}_{DF}}{(1+(n/n_0)^2)^2}
\eea
where $ 	\mathcal{L}_{DF}=\frac{v_F^2\mathcal{H}\tau}{T^2\sigma_Q}, \quad	n_0^2=\frac{\mathcal{H}\sigma_Q}{e^2v_F^2\tau} $. 
Holographic theory with one current gives the same type formula,
\be
L/L_{0}=\frac{A}{(1+(n/n_{0})^{2})^{2}} . 
\ee
\begin{figure}[h]
\begin{center}
{\includegraphics[angle=0,width=0.5\textwidth]{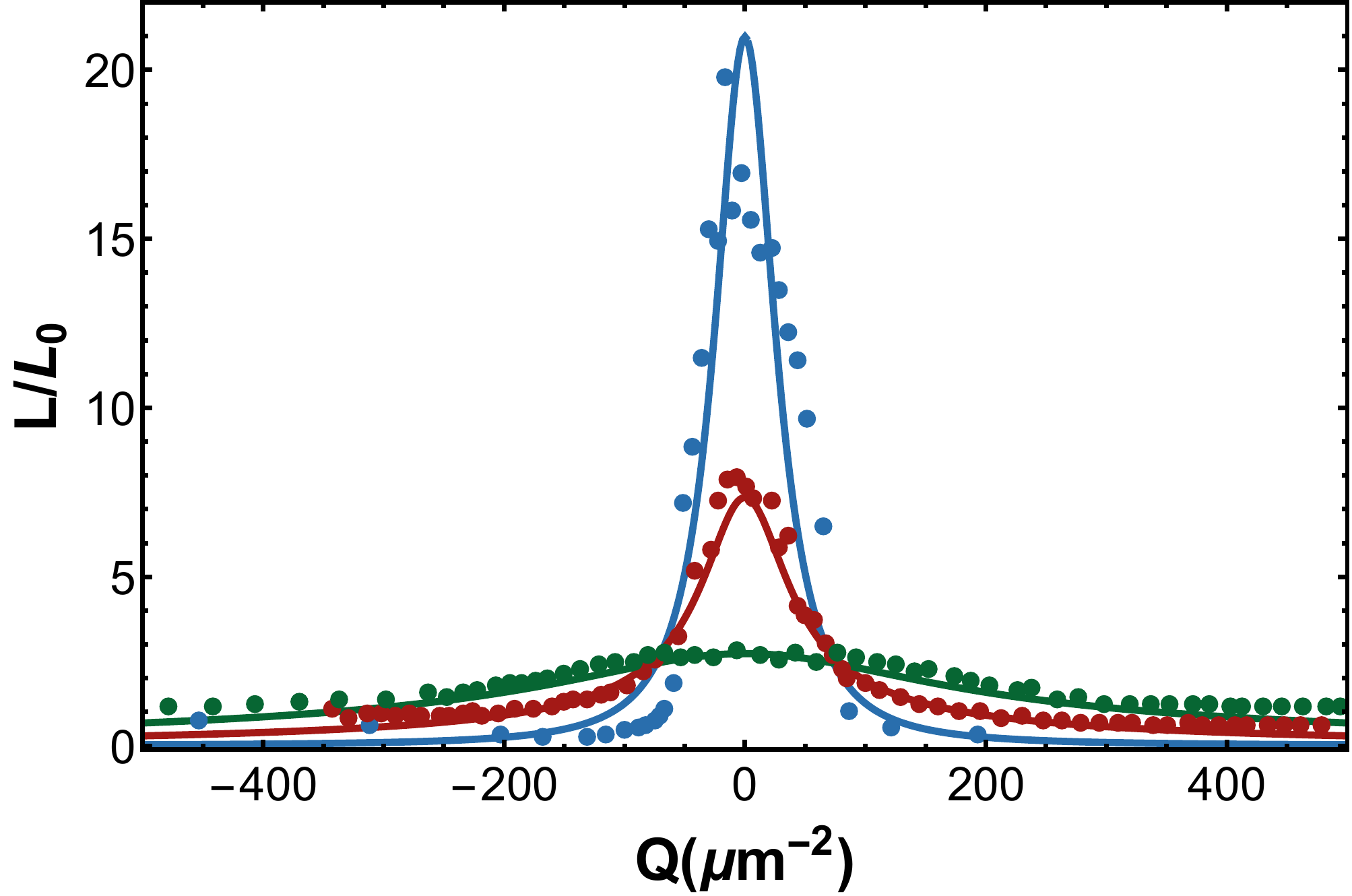}}
 \caption{Lorentz ratio fit by our theory.  Due to the limited temperature dependence of the horizon radius, the holographic theory of $z=1, \theta=0 $ can not fit the data, while that of   $z=3/2, \theta=1 $ theory can fit. 
   } \label{Lorentz} 
\end{center}
\end{figure}
Therefore,   we might   think that the fitting  is trivial, because we have   one more parameter $g_{n}$ with the two current model.  The problem is that  $A\sim s/\sigma_{0}\beta^{2}$ and $n_{0}^{2}\sim s \sigma_{0}\beta^{2}$ and our $\beta$ does not have any $T$ dependence unlike $\tau$ in hydrodynamic theory.  Furthermore, the entropy density  has a specific $r_{0}$ dependence $s\sim r_{0}^{z-\theta}$, which is very sensitive to $z,\theta$. If we treat  $r_{0}$ as a function of other variables as we treat in this paper,  not all the values of $A$ and  $n_{0}$ are available in the necessary  temperature range if  $z=1, \theta=0$.   So the holographic theory has a more tight constraint than the hydrodynamic theory,  as it should be. 
The conclusion is that $z=1, \theta=0$ can not fit the data even with one more parameter $g_{n}$ for the theory with two currents, while  the $z=3/2, \theta=1$ theory can fit.  
See the figure \ref{Lorentz}.

\section{Discussion} We determined the QCP's dynamical exponents for the graphene system. 
The $\theta=1$ result can be interpreted as taking into account the  fermionic nature of the electrons,  and $z=3/2$ can be interpreted as  the flattened  band by the strong interaction.
The $\theta$ was originally introduced to explain the difference of the power in the heat capacity (and entropy density)  between  the bosonic and the fermionic systems. For a boson, $C_{v}\sim T^{d}$ while it is $\sim T$ for fermions.   So, typically $\theta=d-1$ for a fermion system if we define the theta by $C_{v}\sim T^{(d-\theta)}$. In the case of $z\neq 1$, it can be more subtle. At this moment, 
our assumption $\theta=1$ even in the presence of $z\neq 1$ is justified only by data fitting.

\begin{figure}[h]
\begin{center}
\subfigure[$\sigma(T)$]{\includegraphics[angle=0,width=0.4\textwidth]{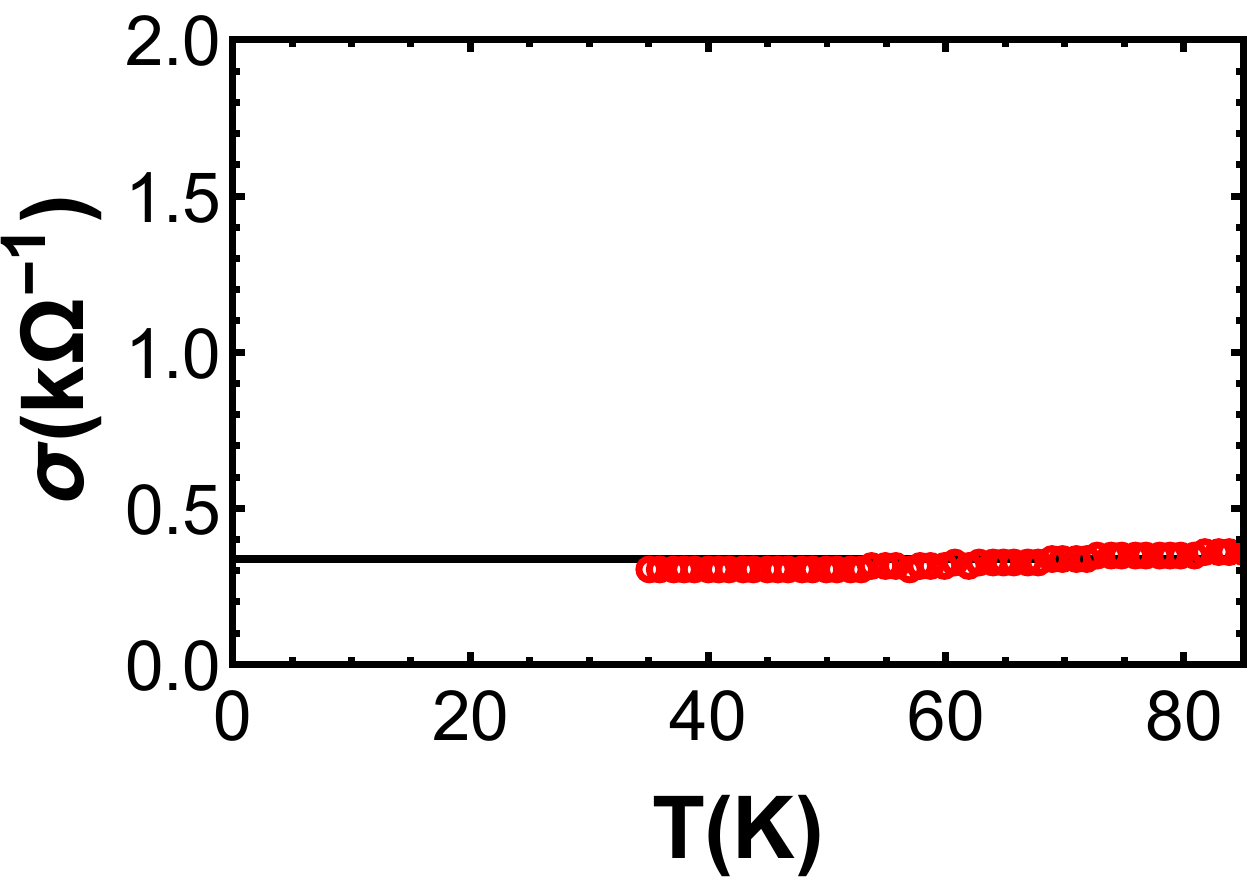}}
\subfigure[$\kappa(T)$]{\includegraphics[angle=0,width=0.4\textwidth]{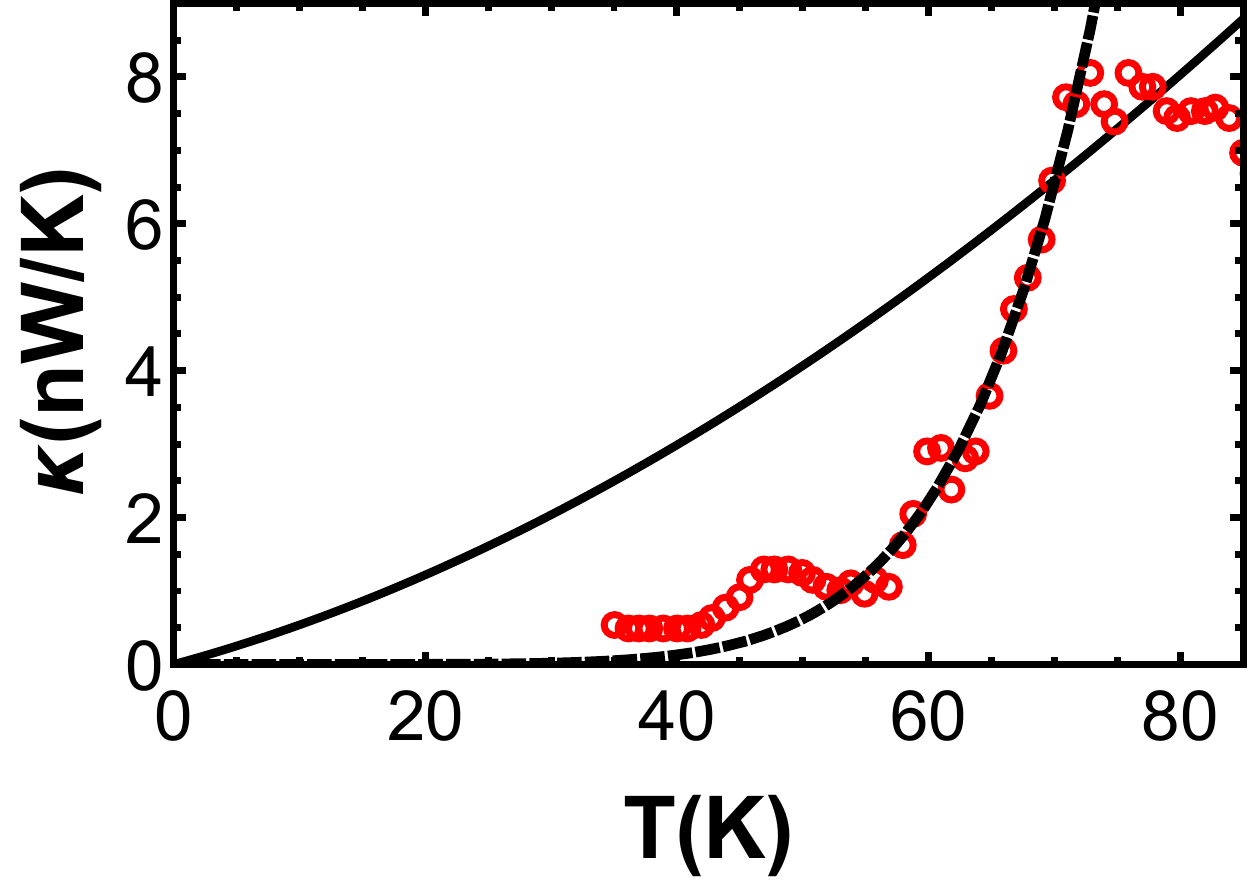}}
\caption{(a) There is not much change in $\sigma(T)$. Red circles are the experimental data, and the solid line is for our model (b).  The fitting by our model (solid line) is not good for varying temperature data. The simple fitting with $T^7$ (dashed curve) misses all the flat feature but can fit rising data quite well. However, we can not explain the power 7 for any reasonable scenario. 
   } \label{fig0} 
\end{center}
\end{figure}

There are some data we can not fit. Since our purpose is to see what kind of data can be fit by holographic theories, we also report such data here because nonfitting is also a  record worthwhile.  
In \cite{Lucas:2015sya}, the authors also computed $\sigma, \kappa$ as functions of $T$.
 Our result to fit the data is given in Fig. \ref{fig0}. 
Figure caption explains how much our theory  does not fit the data. 
Naive scaling law trial that fits the data discarding all flattened part is $\kappa\sim T^{7}$, which can not be explained by any reasonable physics. 
From our experience, we can tell that the density dependence at any fixed temperature could be    fit well, but no temperature dependence at a fixed density is fitted well by holographic  theory. 
We can roughly understand why this is so: 
the data of varying temperatures have   too many features, and it is too steep near 70K; hence, it shows highly unscaling. Therefore, it  can not be fit by any theory that respects the hydrodynamic principle, like a holographic theory: any holographic theory with a horizon should follow hydrodynamics\cite{Bhattacharyya:2008jc}, and therefore, can not fit such data. In fact the data show that as $T$ increases, the minumum values of the carrier density changes drastically. See Fig. \ref{fig2}(a) in~\cite{pkim}.  
The manipulation of the experiment with varying  temperature with fixed density is very hard, because the temperature  can pump up extra charges in a semimetal or a semi-conductor.  
On the other hand, changing the charge density is just matter of changing the chemical potential so varying density data are much more reliable. 
The reason why Ref. \cite{Lucas:2015sya} could 
fit the temperature plot data even roughly  is because the authors cleverly introduced  an idea that  breaks the scale symmetry, the inhomogeneity in the viscosity,  and the coefficient of the derivative expansion. Such an introduction    violates the hydrodynamic principle but is a good choice  for fitting nonscaling data. 
In short, we think that the data of temperature plot are not that of  near QCP in the yellow region of Fig. \ref{region}(a).    
So it is not a surprise for our theory not to fit the temperature plots near the QCP.


\acknowledgments
 This  work is supported by Mid-career Researcher Program through the National Research Foundation of Korea Grant No. NRF-2016R1A2B3007687.  YS is  supported  by Basic Science Research Program through NRF Grant No. NRF-2019R1I1A1A01057998.

\appendix

\section{VALIDITY REGION OF EXPONENTS}
\begin{figure}[h]
\centering
   {\includegraphics[width=5cm]{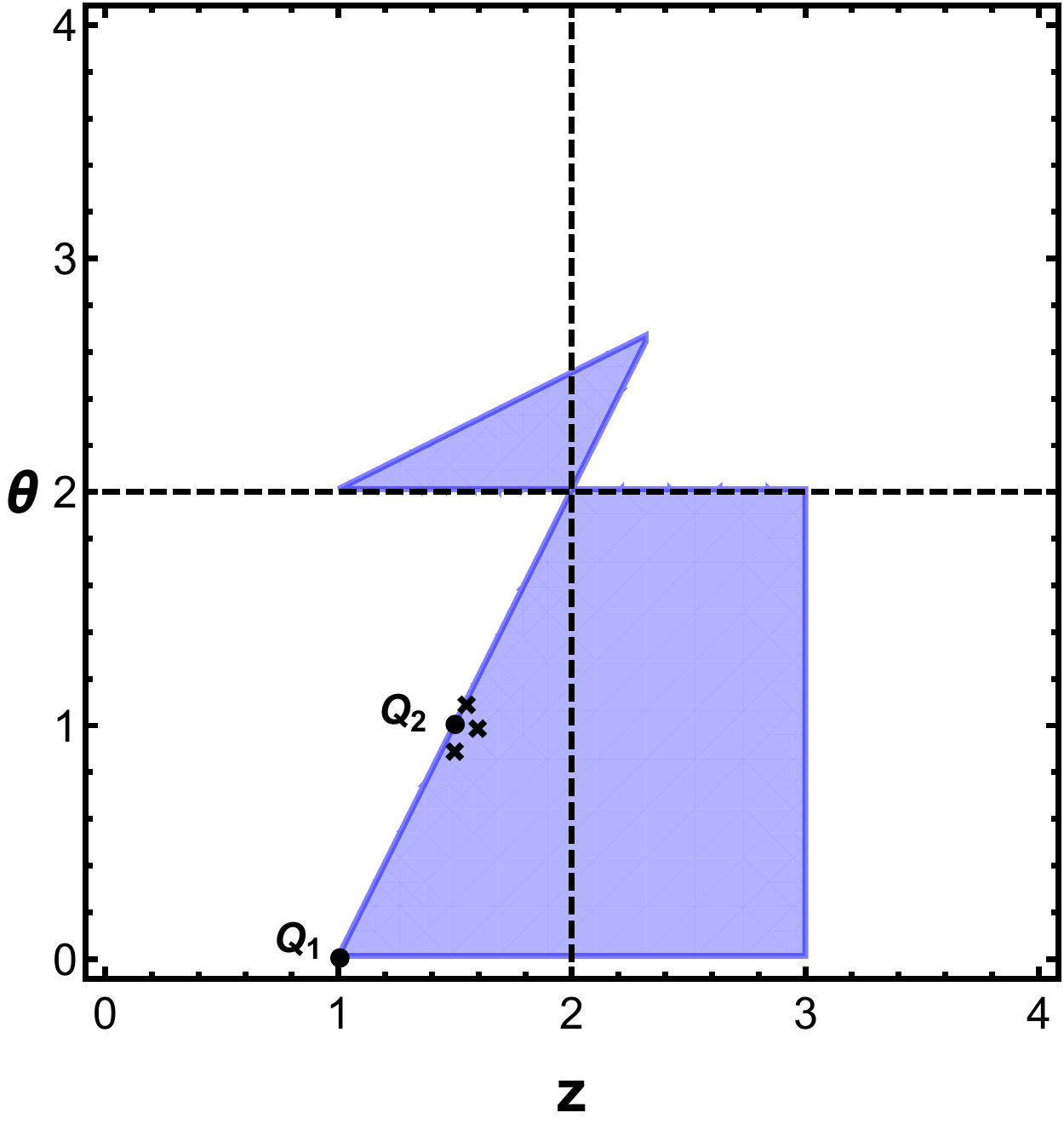}  }
    \caption{   Allowed region of $z,\theta$ coming from null energy condition and positivity of $q_{A}^{2}$.   }
    \label{fig:TI}
\end{figure}

The allowed region for them is described in our previous work\cite{Ge:2019fnj} where we show  that  as a consequence of the null energy condition and positivity of $q_{A}^{2}$, the allowed region are given by the figure \ref{fig:TI}. 

\section{THE RESOLVING SINGULARITY OF HSV GEOMETRY}
In this section, we discuss the curvature singularity in asymptotic hyperscaling violating geometry. We work in a coordinate system that we defined in $(\ref{bgsol})$,
\be
ds^{2}=r^{-\theta}\left(-r^{2z}f(r)dt^{2}+\frac{dr^{2}}{r^2 f(r)}+r^{2}(dx^{2}+dy^{2})\right),\nonumber
\ee
Consider a radial timelike geodesic with four velocity $u=(\dot{t},\dot{r},0,0)$, where the dot denotes $d/d\tau$. The conserved energy is given by $E\equiv-g_{tt}\dot{t}= r^{2z}f(r)\dot{t}$; the normalization of four velocity $u^{\mu}u_{\mu}=-1$ gives
\be
	\dot{r}^2= E^2r^{2(1-z+\theta)}\left(1-\frac{r^{2z-\theta}f(r)}{E^2}\right)
\ee
 Then, we can choose an orthonormal frame for the radial infall of a test particle with an energy $E$, which is given by
 \bea
 	e_0 &&=\frac{E}{r^{2z-\theta}f(r)}\partial_{t} -  E r^{1-z+\theta}\sqrt{1-\frac{r^{2z-\theta}f(r)}{E^2}} \partial_{r} \nonumber\\
 	e_1&&=\frac{E}{r^{2z-\theta}f(r)}\sqrt{1-\frac{r^{2z-\theta}f(r)}{E^2}} \partial_{t} -  E r^{1-z+\theta} \partial_{r}\nonumber\\
 	e_i&&=\frac{1}{r^{1-\theta/2}}\partial_{i} . \eea
Using these basis, we can have the Riemann curvature tensor in this orthonormal frame:
\bea
	R_{abcd}=R_{\mu\nu\rho\sigma}(e_a)^{\mu}(e_b)^{\nu}(e_c)^{\rho}(e_d)^{\sigma}
\eea
The nonvanishing components of the Riemann curvature tensor at the horizon are given by
\bea
	R_{0101} &&= \frac{1}{2}\rh^{1+\theta}\left(  (1+3z-\theta)f'(\rh)+\rh f''(\rh)\right)\nonumber\\
	R_{0i0i}&&=R_{1i1i}\nonumber\\
	&&= \frac{1}{4}\rh^{-2z+2\theta}(\theta-2)\left(E^2 (2-2z+\theta)-\rh^{1+2z-\theta}f'(\rh)	\right)\nonumber\\
	R_{0i1i}&&=\frac{1}{4}E^2\rh^{-2z+2\theta}(\theta-2)(2-2z+\theta)
\eea
Hence, the tidal forces at the horizon are always regular.
\bibliography{Refs_2J.bib}

\end{document}